\documentclass[%
twocolumn,
 amsmath,amssymb,
 aps,
pra,
]{revtex4-1}
\usepackage{CJK}
\usepackage{svg}

\usepackage{graphicx}
\usepackage{dcolumn}
\usepackage{comment}
\usepackage{xcolor}
\usepackage[colorinlistoftodos]{todonotes}
\usepackage[colorlinks=true,citecolor=blue,urlcolor=blue]{hyperref}
\usepackage[english]{babel}
\usepackage{bm}
\usepackage{amsmath}
\usepackage{amssymb}
\usepackage{comment}
\usepackage{subfigure}
\usepackage{epstopdf}
\usepackage{appendix}
\usepackage{float}
\usepackage[normalem]{ulem}
\usepackage{natbib}

\usepackage{mathtools}
\usepackage{physics}
\DeclareUnicodeCharacter{03B2}{\ensuremath{\beta}}
\UseRawInputEncoding

\begin{document}

\title{Three-Body Recombination of Ultracold Microwave-Shielded Polar Molecules}

\author{Ian Stevenson$^{1}$}
\thanks{These authors contributed equally.}
\author{Shayamal Singh$^{2}$}
\thanks{These authors contributed equally.}
\author{Ahmed Elkamshishy$^{2}$}
\author{Niccol\'o Bigagli$^{1}$}
\author{Weijun Yuan$^{1}$}
\author{Siwei Zhang$^{1}$}
\author{Chris~H.~Greene$^{2,3}$}
\email{chgreene@purdue.edu}
\author{Sebastian Will$^{1}$}
\email{sebastian.will@columbia.edu}

\affiliation{$^{1}$Department of Physics, Columbia University, New York, NY, USA}
\affiliation{$^{2}$Department of Physics and Astronomy, Purdue University, West Lafayette, Indiana 47907, USA }
\affiliation{$^{3}$Purdue Quantum Science and Engineering Institute, Purdue University, West Lafayette, Indiana 47907 USA}

\date{\today}

\begin{abstract}
A combined experimental and theoretical study is carried out on the three-body recombination process in a gas of microwave-shielded polar molecules. For ground-state polar molecules dressed with a strong microwave field, field-linked bound states can appear in the intermolecular potential. We model three-body recombination into such bound states using classical trajectory calculations. Our results show that recombination can explain the enhanced loss rates observed at small microwave detunings in trapped samples of bosonic NaCs [Bigagli, \textit{et al.}, Nat. Phys. \textbf{19} 1579-1584 (2023)]. Specifically, our calculations reproduce the experimentally measured three-body loss rates across a wide range of microwave Rabi couplings, detunings, and temperatures. This work suggests that for bosonic shielded molecular systems in which the two-body loss is sufficiently suppressed and a field-linked bound state is present, the dominant loss process will be three-body recombination.
\end{abstract}

\maketitle


Three body-recombination is a fundamental process in which three particles collide, resulting in the formation of a two-body bound state while the third particle allows for conservation of energy and momentum.  Recombination has been studied in systems as varied as anti-hydrogen~\cite{robicheaux2004three} and atmospheric ozone production~\cite{mirahmadi2022three}.  For ultracold atomic systems, recombination provides the fundamental limit on achievable densities and efficiency of evaporative cooling. As such, ultracold atomic recombination has been studied in detail~\cite{hess1983observation, burt1997coherence, esry1999recombination} and is inseparably linked to the development of magnetic Feshbach resonances as a tool in ultracold atomic systems~\cite{chin2010feshbach}.  Seminal work on the three-body problem in ultracold systems has included observations of Efimov trimers \cite{efimov1970energy, efimov1971weakly, kraemer2006evidence}.  For ultracold molecular systems, there have been no studies on three-body recombination.  Instead, losses in molecular systems have been dominated by inelastic two-body collisions \cite{ospelkaus2010quantum, bause2023ultracold}.  However, recent advances in microwave shielding, where microwave fields are used to engineer repulsive barriers in the intermolecular potential \cite{gorshkov2008suppression, cooper2009stable, karman2018microwave, lassabliere2018controlling}, have suppressed the two-body loss \cite{anderegg2021observation, schindewolf2022evaporation, bigagli2023collisionally, lin2023microwave}. Further, measurements in shielded molecules see significant disagreement with the two-body theory \cite{bigagli2023collisionally, lin2023microwave} and it has been proposed that three-body recombination is the reason \cite{bigagli2023observation}.

\begin{figure*}[t]
    \includegraphics[width=\textwidth]{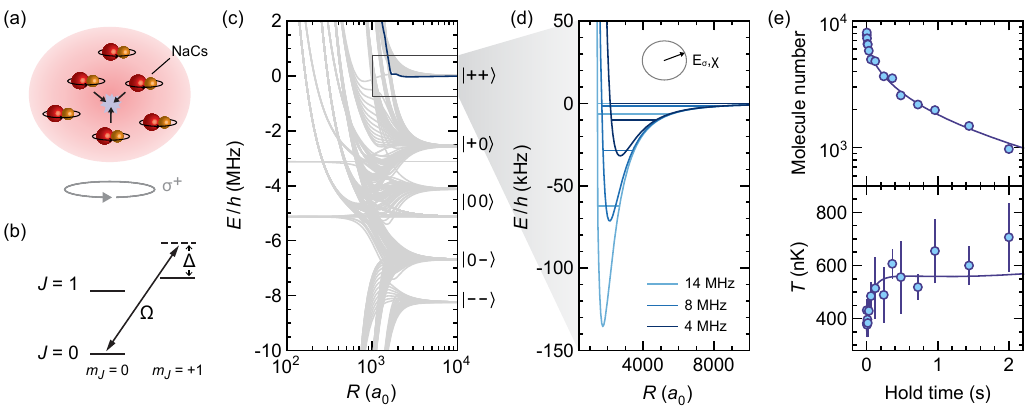}\\
    \caption{Semi-classical three-body loss in microwave-shielded molecules. (a) Shielded NaCs molecules are observed to undergo three-body loss when held in an optical dipole trap. (b) Partial energy level diagram for microwave shielding. Molecules are dressed by a $\sigma^+$ field with Rabi coupling $\Omega$ and detuning $\Delta$. Here, $J$ is the total angular momentum of the molecule and $m_J$ is its projection onto the quantization axis defined by the 864~G magnetic field. (c) Calculated potential energy curves for the collision fo two microwave-shielded NaCs molecules for $\Omega / (2 \pi) = 4$~MHz and $\Delta / \Omega = 0.25$. The blue line shows the $L = 0$, $\ket{++}$ shielded potential. Potentials in grey either correspond to higher orbital angular momentum or unshielded states. (d) $L = 0$ adiabatic potential energy curves on the kHz scale for different $\Omega$ and fixed $\Delta / \Omega = 0.25$. The horizontal lines show the bound states in the potentials. The calculation is performed for $\chi = 3(1)$ degree ellipticity of the microwave field (illustrated in inset). (e) Measured three-body loss of molecules in an optical dipole trap for a sample that starts with $8(1) \times 10^3$ molecules at $T = 350(50)$~nK. Molecule number is recorded as a function of hold time and fit to a kinetic model, see Supplemental Material~\cite{SI}, to extract the three-body loss rate coefficient, $L_{\rm 3B}$. Data is shown for $\Omega / (2 \pi) = 8$~MHz and $\Delta / \Omega = 0.25$~MHz. The error bars show the standard error of the mean for two repetitions of the experiment. }
    \label{fig:1}
\end{figure*}

In microwave shielding, molecules are dressed by a blue-detuned circularly-polarized $\sigma^+$ microwave field, see Fig.~\ref{fig:1}(a) and (b). Unlike atomic systems where van der Waals interactions dominate, shielded molecules primarily interact through long-range electric dipole-dipole interactions: $V_{\textrm{dd}} = d_{\rm eff}^2 (3 \cos^2{\theta} -1 )/ (4 \pi \epsilon_0 R^3)$. Here, $R$ is the distance between molecules, $\theta$ is the angle between the dipoles and the intermolecular axis and $d_{\rm eff}$ is the induced dipole moment. For shielded molecules, the induced dipole moment is $d_{\rm eff} = d_0 / \sqrt{12 + 12 (\Delta / \Omega)^2}$, where $d_0$ is the permanent dipole moment of the molecule, $\Delta$ is the microwave detuning and $\Omega$ is the Rabi coupling. Modeling the anisotropic nature of the dipole-dipole interaction adds complexity to the notoriously difficult three-body problem. As such, \textit{ab initio} studies of dipolar three-body recombination are sparse. Early theoretical work has predicted that the recombination rate should scale with $d_{\rm eff}^8$ (under a quantum treatment)~\cite{ticknor2010three} and that the Efimov effect should survive for dipolar systems~\cite{wang2011efimov}. 

For microwave-shielded NaCs molecules, collisions are semi-classical. The typical temperatures in the experiment $\sim 100$~nK, are much larger than the characteristic energy scale of the dipolar interactions~\cite{bohn2009quasi, SI}, $E_{\rm d} = 8 (4 \pi \epsilon_0)^2 \hbar^6 / (m^3 d_{\rm eff}^4) \sim 1$~nK ($m$ is the mass of the molecule). For semi-classical collisions, a large number of partial waves contribute to the scattering and thus the $s$-wave scattering length, $a_s$, is not a relevant parameter. Additionally, the scaling laws change. In ultracold atomic recombination, the recombination rate is temperature independent and $L_{\rm 3B} \propto a_s^4$~\cite{fedichev1996three}, where $L_{\rm 3B}$ is the three-body loss rate coefficient. For semi-classical dipolar interactions, the prediction is $L_{\rm 3B} \propto T^{-7/6}d_{\rm eff}^{10/3}$ \cite{SI}. Finally, and critically for our study, we can build on previous work in modeling recombination classically~\cite{perez2014comparison, krukow2016energy}.

This letter reports on a joint theoretical and experimental study of three-body recombination in microwave-shielded NaCs molecules. Although the microwave-induced barrier blocks recombination into short-range bound states, there can exist long-range field-linked bound states~\cite{avdeenkov2003linking, chen2023field} outside of the microwave barrier, as shown in Fig.~\ref{fig:1}(c) and (d). The experiment measures three-body loss in ensembles of microwave-shielded NaCs molecules. Using classical trajectory methods we calculate rate coefficients for recombination into the field-linked bound states as a function of $\Delta$, $\Omega$, and sample temperature, $T$. We find excellent quantitative agreement between the calculated recombination rate and the experimental three-body loss rate. This agreement supports the hypothesis proposed in Ref.~\cite{bigagli2023observation}: the presence of bound states in the shielded collisional potential gives rise to three-body recombination as the dominant loss process. 


First, we turn our attention to measuring the three-body loss rate for shielded molecules. Our experiment begins with a gas of $10^4$ NaCs molecules at $250(50)$~nK, collisionally shielded by a $\sigma^+$-polarized microwave field. The molecules are held in a crossed dipole trap with trap frequencies $2\pi \times \{ 50(5), \ 50(5), \ 140(5)\}$~Hz and $3.6(4)$~$\mu$K trap depth. More detail on the preparation and detection of the molecules may be found in Refs.~\cite{stevenson2023ultracold, warner2023efficient, bigagli2023collisionally}. We hold the molecules in the trap and measure the evolution of the number, $N$, and temperature, $T$, as shown in Fig.~\ref{fig:1}(e). In parallel work, we show that we can clearly identify the loss process as three-body~\cite{yuan2024upcoming}. The data is fit to a kinetic model~\cite{SI} to extract $L_{\rm 3B}$. We perform the experiments for $\Omega / (2 \pi) = 4.0(1), 8.0(2), 14.0(5)$ and $31(1) $~MHz and vary the detuning in a range corresponding to $0 < \Delta / \Omega < 2$. This will allow us to evaluate the scaling of $L_{\rm 3B}$ with $d_{\rm eff}$. Finally, to elicit the temperature dependence of $L_{\rm 3B}$, we can evaporate to a range of target temperatures and then record the loss rate coefficient.


\begin{figure} [t]
\centering
    \includegraphics[width=\linewidth]{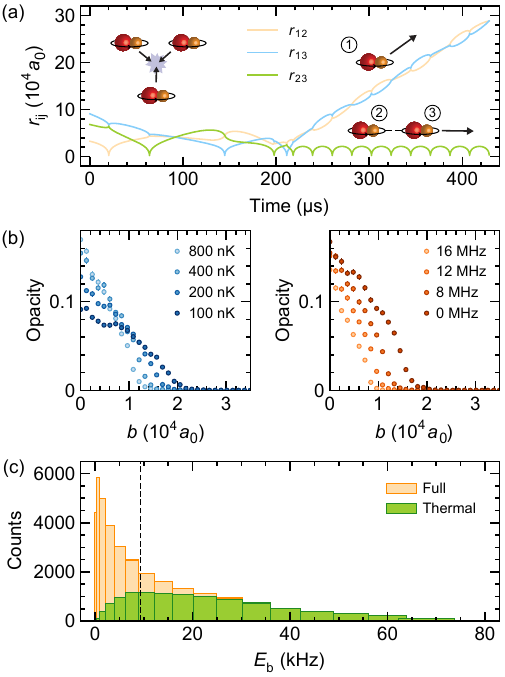}\\
    \caption{Classical trajectory methods applied to microwave-shielded NaCs molecules. (a) Example of a recombination trajectory for $E=200\,\textrm{nK}$ and $b=0\,a_0$. As a result of the collision, molecules $2$ and $3$ recombine into a bound state and molecule $1$ flies out with excess energy. (b) Calculated opacity function. (right panel) $\Tilde{\mathcal{P}}(b,E,T=300~{\rm nK})\{8,0\}$ as a function of impact parameter, $b$, for a range of collisional energies. There is a maximum impact parameter for a given energy beyond which the opacity function goes to zero. (left panel) $\Tilde{\mathcal{P}}(b, E=500~{\rm nK}, T = 300~{\rm nK})\{8,\Delta\}$ as a function of $b$, for $\Delta$ ranging from 0 to 16~MHz. (c) Calculated thermal dissociation of tetramers. The orange histogram shows all trajectories resulting in bound molecules, plotted as a function of their binding energy. The green histogram shows the relevant trajectories after accounting for thermal dissociation. Calculation is done for $\Omega = 2 \pi \times 8$~MHz and $\Delta/\Omega = 1$ at a collision energy of $k_{\rm B} \times 400$~nK and a sample temperature of $300$~nK. The black dashed line marks the binding energy of the quantum bound state. }
    \label{fig:2}
\end{figure}

Next, we consider the problem of theoretically calculating the recombination rate coefficient. Building on previous work in modeling three-body recombination classically \cite{perez2014comparison}, the problem can be reduced to estimating the opacity function $\Tilde{\mathcal{P}}(b,E,T)\{ \Omega,\Delta \}$. The opacity function represents the fraction of three-body collisions that result in bound molecules, as a function of the microwave parameters $\{ \Omega,\Delta \}$, the impact parameter $b$, the initial energy $E$ and the sample temperature $T$. An example of a recombination event is shown in Fig.~\ref{fig:2}(a) and an example of the opacity function is shown in Fig.~\ref{fig:2}(b). To estimate the opacity function, $10^4$ Newtonian trajectories are computed for each combination of parameters. When counting bound trajectories, we check if the tetramers can thermally dissociate as part of a collision with another molecule. Such tetramers would not be observed as recombination in an experiment. As can be seen in Fig.~\ref{fig:2}(c), thermal dissociation has a profound effect on which collisions lead to tetramer formation: most of the weakly bound tetramers with binding energies less than the temperature are broken up again. More details on including thermal dissociation can be found in the supplementary material~\cite{SI}. From the opacity function we can recover the temperature-dependent recombination rate, $K_3(T)$, by integrating over the impact parameter and averaging over the initial energy weighted by the Maxwell-Boltzmann distribution~\cite{perez2014comparison, SI}.

We calculate the classical motion of three molecules starting from the three-body Hamiltonian:
\begin{equation}
\label{eq:1}
H =  \frac{\Vec{p_1}^2}{2\mu_{12}}+\frac{\Vec{p_2}^2}{2\mu_{3,12}} + V(\vec{r}_1, \vec{r}_2, \vec{r}_3) .
\end{equation}
For equal masses, $\mu_{12}=m/2$ and $\mu_{3,12}=2m/3$. $\Vec{p_i}$ is the momentum conjugate to the Jacobi vector $\Vec{\rho_i}$ where $\Vec{\rho}_1=\Vec{r}_1-\Vec{r}_2$ and $\Vec{\rho}_2=\Vec{r}_3-(\Vec{r}_1+\Vec{r}_2)/2$. It should be noted that the momentum terms in Eq.~\ref{eq:1} treat angular momentum classically. Following the treatment of Ref.~\cite{perez2014comparison}, we estimate the three-body potential as the pairwise sum of reduced two-body potentials:
\begin{gather}
V(\vec{r}_1, \vec{r}_2, \vec{r}_3) = \Tilde{U}_{++}(r_{12})+\Tilde{U}_{++}(r_{23})+\Tilde{U}_{++}(r_{31}).
\end{gather}
For microwave-shielded molecules, where the molecules are dressed by a blue-detuned $\sigma^+$ polarized microwave field, the two-body adiabatic potentials are obtained by diagonalizing the two-body Hamiltonian given by \cite{karman2018microwave}:
\begin{multline} \label{eq:4}
    \hat{H} = \hat{h}_1 + \hat{h}_2 + \hat{h}_{mw} -\Vec{E}\cdot(\vec{d_1}+\vec{d_2} ) + \frac{\vec{L}^2}{2\mu R^2} + \hat{V}_{\rm dd}.
\end{multline}
Here, the first two terms describe the rotation of each dipole separately, $\hat{h}_i = B\Vec{J_i^2}$; the third term describes the free field Hamiltonian and can be written in terms of photon creation and annihilation operators, $  \hat{h}_{mw} = \hbar \omega \hat{a}_{\omega}^{\dag}\hat{a}_{\omega}$; the fourth term is the field-dipole interaction, $-\Vec{E}\cdot\Vec{d} = -E_0[({\vec{d}}\cdot \hat{\eta})\hat{a}_{\omega}^{\dag} + h.c.] / \sqrt{N_0}$ ($N_0$ is the number of photons in a proper quantization volume); the centrifugal term, $\vec{L}^2 / (2\mu R^2)$, describes the rotation of the two dipoles around their center of mass ($\vec{L}^2$ is the relative orbital angular momentum of the two dipoles); the final term, $\hat{V}_{\rm dd}$, is the dipole-dipole interaction. Diagonalizing Eq.~\ref{eq:4} with the centrifugal term averages over the dipolar anisotropy but mixes $L = 2$ character into the lowest shielded potential, $U_{++}(r_{ij})$. As to not double count the angular momentum, we subtract the expectation value of $\Vec{L}^2/2\mu R^2$ from the potentials, $\Tilde{U}_{++}(r_{ij})=U_{++}(r_{ij})-\left<\Vec{L}^2/2\mu_{ij}r_{ij}^2\right>_{++}$ for the classical scattering calculations using Eq.~\ref{eq:1}. We note that the calculation of the two-body adiabatic potentials takes into account the finite ellipticity of the circular microwave field in the experiment. Non-zero ellipticity increases the depth of the potential and thereby the recombination rate.

\begin{figure}[t]
    \includegraphics[width=8.6 cm]{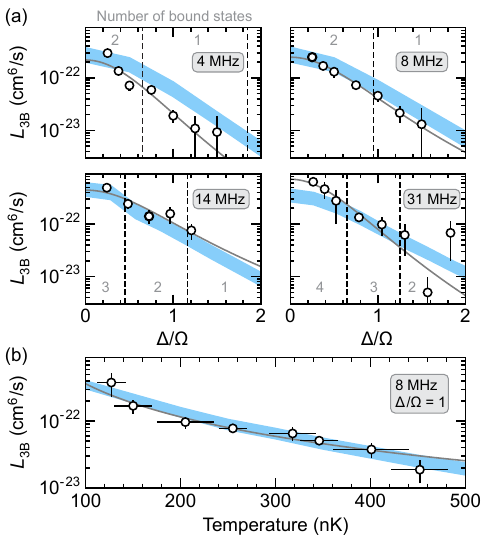}\\
    \caption{Comparison of measured three-body loss rates and calculated recombination rates. (a) $L_{\rm 3B}$ as a function of $\Delta/\Omega$ for $\Omega / (2 \pi) = 4$, 8, 14, and 31~MHz. Circles are the experimental values with error bars representing the error from the fit. Blue-shaded bands are the result of the trajectory calculation. The width of the bands reflects the experimental uncertainty of the ellipticity, 3(1) degrees. Vertical dashed lines indicate the appearance of a bound state in the collisional potential, grey numbers denote the number of bound states in the potential adiabatic to $L=0$. Solid grey lines are fits to $L_{\rm 3B} = a d_{\rm eff}^x$, yielding $x = 6.8(1.3), \ 5.2(0.3), \ 4.2(1.1), \ 6.3(1.0)$ for $\Omega / (2 \pi) = 4$, 8, 14, and 31~MHz respectively. (b) $L_{\rm 3B}$ as a function of temperature for  $\Omega = 2 \pi \times 8$~MHz and $\Delta/\Omega = 1$. Grey solid line shows a fit to $L_{\rm 3B} = a T^y$ yielding $y = -1.7(2)$. 
    }
    \label{fig:3}
\end{figure}


The comparison of the experimental data with the classical model is shown in Fig.~\ref{fig:3}. The key assumption in this comparison is that the tetramer formed in the recombination process undergoes a fast loss process. An important distinction is between the recombination rate $K_3$, which is calculated in the theory \cite{greene2017universal}, and the three-body loss rate $L_{\rm 3B}$, which is experimentally relevant and needs to take into account the number of molecules lost in a recombination event. When the binding energy is larger than the trap depth, all three molecules are lost and $L_{\rm 3B} \approx K_3 / 2$. In the experiment this is only the case for the data at $\Omega/ (2 \pi)=31$~MHz and the smallest detunings at $\Omega/ (2 \pi)=14$~MHz. For all other data, the trap is sufficiently deep such that only the tetramer is lost, yielding $L_{\rm 3B}=K_3/3$. For $\Omega / (2 \pi) =8$, 14, and 31~MHz the agreement between the data and calculation is excellent. For $\Omega / (2 \pi) =4$~MHz the agreement is less good. It is unclear if this is a systematic effect or indicative of a more fundamental change in the collisional physics. The blue-shaded uncertainty regions in Fig.~\ref{fig:3}, which reflect the $\pm 1$ degree uncertainty of the microwave ellipticity, indicate that small deviations of the ellipticity can have a significant impact on the loss rate. Finally, we note that, consistent with semi-classical scattering, we observe no resonances associated with bound states entering the potential.

Next, we examine the agreement between the experimental data, the trajectory model, and the expected scaling law, $L_{\rm 3B} \propto T^{-7/6} d_{\rm eff}^{10/3}$. We note that before accounting for thermal dissociation, the trajectory model yields the same temperature scaling as the scaling law, while the scaling with dipole moment, $d_{\rm eff}^x$, yields $x = 4.2(0.2)$~\cite{SI}. Presumably this different value of the exponent reflects the significant deviation of the interaction potential at short-range from a pure dipole interaction. First, we use our measurement of $L_{\rm 3B}$ versus $\Delta/\Omega$, Fig.~\ref{fig:3}(a), to examine the scaling with $d_{\rm eff}$. Averaging the results from the four experimental data sets, we obtain $x = 5.3(0.6)$. Notably, the power law fits are pulled to larger values because of the first point, $\Delta / \Omega \approx 0.25$. Excluding these points, $x = 4.8(0.7)$. Perhaps, higher body effects, like four-body loss, become relevant when the detuning is very small. For comparison, the trajectory model still follows $x = 4.2(0.2)$ when thermal dissociation is taken into account, in agreement with the experimental data. Second, we examine the scaling of $L_{\rm 3B}$ with temperature. The results for $\Omega / (2 \pi) = 8$~MHz, $\Delta / \Omega = 1$ are shown in Fig.~\ref{fig:3}(b). Fitting a power law $L_{\rm 3B} \propto T^y$ to the experimental data, we extract the exponent $y = -1.7(2)$. While both deviate from the scaling law, the experiment and trajectory model are in excellent agreement.


The excellent agreement between the trajectory model and the experiment suggests that recombination in conjunction with a fast loss process for (NaCs)$_2$ tetramers can explain the observed three-body loss rates. A likely loss process for (NaCs)$_2$ tetramers are collisions with the unpaired molecules. The loss cannot be explained by a potential expulsion of tetramers from the trap due to the energy release of recombination events, as the binding energy of the bound states is small, often less than 10~kHz. The assumption of fast loss of the tetramers is in contrast to the recent finding in Ref.~\cite{chen2024ultracold}, where field-linked (NaK)$_2$ tetramers formed in a system of fermionic NaK molecules were found to be relatively long-lived. In fact, preliminary calculations suggest that the predissociation lifetime of the (NaCs)$_2$ tetramers would be longer than the timescales of our experiments. However, collisional tetramer-molecule processes may be significantly enhanced for bosonic NaCs compared to fermionic NaK. A similar effect is known from atomic samples close to a Feshbach resonance, where weakly bound molecules of fermionic atoms are found to be collisionally stable~\cite{petrov2005scattering}, while their bosonic counterparts have high loss rates from collisional relaxation.

In conclusion, we have shown that three-body recombination into field-linked bound states is the dominant loss mechanism for microwave-shielded NaCs molecules when the microwave detuning is small, i.e.~$\Delta < \Omega$. While our study focuses on NaCs, we expect our findings to be generally relevant for ultracold bosonic molecules, such as NaRb~\cite{lin2023microwave}, CaF~\cite{anderegg2021observation}, SrF~\cite{jorapur2024high}, and YO~\cite{burau2023blue}. The recombination mechanism discussed here is also likely to be relevant for other shielding approaches, as the resulting collisional potentials often come with an attractive well that supports bound states. Besides microwave shielding, this includes resonant shielding with static electric fields~\cite{wang2015tuning, matsuda2020resonant, mukherjee2024controlling} and shielding approaches combining a static field with a microwave field~\cite{gorshkov2008suppression}. Going forward, we expect that there will be a need to identify shielding approaches that allow the engineering of all-repulsive collisional potentials in order to suppress recombination losses, such as double microwave shielding~\cite{bigagli2023observation, karman2024upcoming}. Mitigation of three-body losses is critical, e.g., for evaporative cooling of bosonic molecules into a Bose-Einstein condensate~\cite{bigagli2023observation}.

While we see excellent agreement between the experimental data and the classical model in the temperature regime of a few hundred nanokelvin, the model should become inaccurate for temperatures where the scattering is clearly quantum. For NaCs, this will be the case for temperatures on the order of one nanokelvin. This can potentially be studied with a BEC of NaCs molecules and possibly with degenerate gases of other molecules, such as NaRb and NaK, in the future. It is an interesting question, what specifically will change, when the scattering becomes quantum. In addition, our work suggests that tetramers formed in recombination are quickly lost in tetramer-molecule collisions. Theoretical modeling of such processes and how they differ in detail for bosonic and fermionic molecules will be of great interest.

We thank Tijs Karman for helpful discussions and Jes\'{u}s P\'{e}rez-R\'{i}os for access to computer programs. This work was supported by an NSF CAREER Award (Award No.~1848466), an ONR DURIP Award (Award No.~N00014-21-1-2721), an AFOSR-MURI award (Award No.~FA9550-20-1-0323), and a grant from the Gordon and Betty Moore Foundation (Award No.~GBMF12340). W.Y.~acknowledges support from the Croucher Foundation. I.S.~was supported by the Ernest Kempton Adams Fund. S.W.~acknowledges additional support from the Alfred P.~Sloan Foundation.

\end{document}


\title{Supplemental Material for \\ ``Three-Body Recombination of Ultracold Microwave-Shielded Polar Molecules''}
\author{Ian Stevenson}
\affiliation{Department of Physics, Columbia University, New York, New York 10027, USA}
\author{Shayamal Singh}
\affiliation{Department of Physics and Astronomy, Purdue University, West Lafayette, Indiana 47907, USA }
\author{Ahmed Elkamshishy}
\affiliation{Department of Physics and Astronomy, Purdue University, West Lafayette, Indiana 47907, USA }
\author{Niccol\'o Bigagli}
\affiliation{Department of Physics, Columbia University, New York, New York 10027, USA}
\author{Weijun Yuan}
\affiliation{Department of Physics, Columbia University, New York, New York 10027, USA}
\author{Siwei Zhang}
\affiliation{Department of Physics, Columbia University, New York, New York 10027, USA}
\author{Chris~H.~Greene}
\affiliation{Department of Physics and Astronomy, Purdue University, West Lafayette, Indiana 47907, USA }
\affiliation{Purdue Quantum Science and Engineering Institute, Purdue University, West Lafayette, Indiana 47907 USA}
\author{Sebastian Will}
\affiliation{Department of Physics, Columbia University, New York, New York 10027, USA}

\date{\today}

\maketitle

\section{S1. Quantum two-body calculation}
The two-body interaction Hamiltonian is given by:
\begin{align} 
\label{ad ham}
    \hat{H}_{2B} = \hat{h}_1 + \hat{h}_2 + \hat{h}_{mw} -\Vec{E}\cdot(\vec{d_1}+\vec{d_2} ) + \frac{\vec{L}^2}{2\mu R^2} + \hat{V}_{dd}.
\end{align}
The first two terms describe the rotation of each dipole separately where $\hat{h}_i = B\Vec{J_i^2}$ and $B$ is the rotational constant; the term $
\hat{h}_{mw}$ describes the free field Hamiltonian and can be written in terms of photon creation and annihilation operators as  $  \hat{h}_{mw} = \hbar \omega \hat{a}_{\omega}^{\dag}\hat{a}_{\omega}$; the fourth term is the field-dipole interactions; the term $\frac{\vec{L}^2}{2\mu R^2}$ describes the rotation of the two dipoles around their center of mass; and the final term is the dipole-dipole interaction written as a sum of irreducible products of vector and tensor operators as \cite{varshalovich1988quantum}
\begin{align}
\label{int_pot}
    \hat{V}_{dd} =  d^2\times\sum_{q_2,q_1}  f_{q_1,q_2} 
    \frac{\mathcal{Q}_{1}^{q_1}(\hat{r}_1) \mathcal{Q}_{1}^{q_2}(\hat{r}_2)\mathcal{Q}_{2}^{-(q_1+q_2)}(\hat{R})}{R^{3}},
\end{align}
where 
\begin{equation}\label{mult_momement}
    \mathcal{Q}_{L_A}^M(\hat{r}_A) = \sqrt{\frac{4 \pi}{2L+1}} Y_{L_A}^{M}(\theta_A,\phi_A)
\end{equation} 
and
\begin{equation}
    \label{f_consts}
f_{q_1,q_2} = 
(-1)^{q_1}\binom{2+q_1+q_1}{1+q_2}^{1/2}\binom{2-(q_1+q_2)}{1-q_2}^{1/2}
\end{equation}
and $\vec{L}^2$ is the relative orbital angular momentum of the two dipoles. The field-dipole interaction is written in terms of field operators as 
\begin{equation}
    -\Vec{E}\cdot\Vec{d} = -\frac{E_0}{\sqrt{N_0}}(({\vec{d}}\cdot \hat{\eta}) \hat{a}_{\omega}^{\dag} + h.c.),
\end{equation}
where $N_0$ is the number of photons in a proper quantization volume. Ultimately, the energy will always be shifted by the zero point energy of the field $N_{0} \hbar \omega$. The field-dipole interaction mixes the rotational states of two dipoles. The strength of the electric field defines a shielding radius $R_0$. In this region, the dipole-dipole interaction is comparable to microwave-dipole interactions. At longer distances $R> R_0$, the dipole-dipole interaction becomes comparable to the centrifugal forces of the two-dipole systems. This gives rise to an extremely long-range nature of the two-body adiabatic potentials. The two-body adiabatic potentials are obtained by diagonalizing the Hamiltonian in Eq.~\ref{ad ham} in the basis that represents two uncoupled states of each dipole $(j_i,m_{j_i})$ where $i = 1,2$, a partial wave describing the relative angular momentum of the two dipoles $(L,m_L)$, and a photon state representing the number of photons interacting with the dipoles, $N$.
The basis functions are written explicitly as:
\begin{equation}
    \ket{\phi_k} = {\cal S}(\ket{j_1,m_1 }\ket{j_2,m_2}) \ket{L,m_L} \ket{N},
\end{equation}
\begin{equation}
    \braket{\hat{r}_i}{j_i,m_i} = Y_{j_i,m_i}(\hat{r}_i),
\end{equation}
\begin{equation}
    \braket{\hat{R}}{L,m_L} = Y_{L,m_L}(\hat{R}).
\end{equation}
Here, the index $k$ is a collective index for the seven quantum numbers $(j_1,m_1,j_2,m_2,L,m_L,N)$. The Bose symmetrizer is given by $ {\cal S} = \frac{1+P_{12}}{\sqrt{2(1+\delta_{j_1,j_2}\delta_{m_1,m_2})}}$, and $P_{12}$ is the permutation operator. Since we work with a system of two identical bosons, the values of $L$ are restricted to be even (0,2,4,...).
Next, the matrix element of the adiabatic Hamiltonian is calculated between two basis functions, obtained by separating the Hamiltonian into three parts $H_{ad} = H_1 + H_2 +H_3$:
\begin{equation}
    H_1 = \hat{h}_1 +\hat{h}_2 + \hat{h}_{mw} + \frac{\vec{L}^2}{2\mu R^2},
\end{equation}
\begin{equation}
    H_2 = -\Vec{E}\cdot(\vec{d_1} + \vec{d_2}), 
\end{equation}
\begin{equation}
    H_3 = \hat{V}_{dd}.
\end{equation}
The first part is diagonal and given by 
\begin{align} \label{h1}
    \bra{\phi_{k'}}H_1\ket{\phi_k} = { [ }B(j_1(j_1+1) + j_2(j_2+1)) 
    + N\hbar\omega + \frac{L(L+1)}{2\mu R^2}) {] }\delta_{k',k}.
\end{align}

The matrix element of $H_2$ is computed as
\begin{multline} 
\label{h2}
    \bra{\phi_{k'}}H_2\ket{\phi_k} =  -E_0  \delta_{L, L'} \delta_{M_L, M_L'} 
      \times
       \left( \delta_{j_2', j_2}\delta_{m_2', m_2} \left[ \bra{j_1', m_1'}\hat{d}_{\sigma}\ket{j_1, m_1} \delta_{N', N+1} 
     + \bra{j_1', m_1'}\hat{d}_{\sigma}^{\dagger}\ket{j_1, m_1} \delta_{N', N-1} \right] 
     \right. \\ \quad  \left.
      + \delta_{j_1', j_1}\delta_{m_1', m_1} \left[ \bra{j_2', m_2'}\hat{d}_{\sigma}\ket{j_2, m_2} \delta_{N', N+1} 
      + \bra{j_2', m_2'}\hat{d}_{\sigma}^{\dagger}\ket{j_2, m_2} \delta_{N', N-1} \right] \right) ,
\end{multline}
where $\hat{d}_\sigma = \cos{\gamma} \hat{d}_+ - \sin{\gamma} \hat{d}_{-}$, $\gamma$ is the ellipticity angle of the electric field, and $d_{\pm} = \mp d \times \mathcal{Q}_{1 \pm 1}$. 
The matrix element of the dipole-dipole interaction is directly computed as 
\begin{align} 
\label{h3}
    \bra{\phi_{k'}}H_3\ket{\phi_k} = d^2 \frac{\delta _{N',N}}{R^3} 
    \times \sum_{q_1,q_2}f_{q_1,q_2}\times \bra{j_1', m_1'}Q_{1}^{q_1}\ket{j_1, m_1} \times  \bra{j_2', m_2'}Q_{2}^{q_2}\ket{j_2, m_2}  \bra{L', m_L'}Q_{2}^{-q_1-q2}\ket{L, m_L}
\end{align}  
and finally, the matrix element of a spherical tensor $\mathcal{Q}_L^M$ between two angular momentum states is given in terms of Clebsch-Gordan coefficients by 
\begin{equation} \label{Y3}
    \bra{l', m'} \mathcal{Q}_L^M \ket {l, m }  = \sqrt{\frac{(2L+1)(2l+1)}{2l'+1}}C_{lm,LM}^{l'm'}C_{l0,L0}^{l'0}
\end{equation}

\section{S2. Classical trajectory calculations}
To describe the collision between molecules dressed by the microwave electric fields, the interaction is described as a sum of pairwise potentials from the quantum two-body calculation. It should be noted that the interaction in the quantum calculation is non-central, but we obtain two-body radial potentials by diagonalizing the adiabatic Hamiltonian with $\Vec{L}^2$ included, which averages over the anisotropy. To obtain bare radial potentials that do not include the average angular energy, the expectation value of $\Vec{L}^2/2\mu R^2$ is subtracted from the potentials, and then the angular momentum is treated classically through the Hamiltonian
\begin{equation}
\label{class_ham}
H =  \frac{\Vec{p_1}^2}{2\mu_{12}}+\frac{\Vec{p_2}^2}{2\mu_{3,12}} + V(\Vec{\rho_1},\Vec{\rho_2}),
\end{equation}
where for equal masses, $\mu_{12}=m/2$, $\mu_{3,12}=2m/3$ and $m$ is the mass of the molecule. $\Vec{\rho}_1=\Vec{r}_1-\Vec{r}_2$, $\Vec{\rho}_2=\Vec{r}_3-(\Vec{r}_1+\Vec{r}_2)/2$ and $\Vec{p_i}$ is the momentum conjugate to the Jacobi vector $\Vec{\rho_i}$. The interaction potential is approximated by
\begin{equation}
V(\Vec{\rho_1},\Vec{\rho_2})=\Tilde{U}_{++}(r_{12})+\Tilde{U}_{++}(r_{23})+\Tilde{U}_{++}(r_{31}),
\end{equation}where the two-body potential 
\begin{equation}\Tilde{U}_{++}(r_{ij})=U_{++}(r_{ij})-\left<\Vec{L}^2/2\mu_{ij}r_{ij}^2\right>_{++}
\end{equation} is obtained from the quantum calculation, $<.>_{++}$ represents expectation value in the $\ket{++}$ state and $r_{ij}=|\Vec{r}_i-\Vec{r}_j|$ is the intermolecular distance.
Following the treatment in \cite{perez2014comparison,greene2017universal}, a 6D position vector is constructed with $\Vec{\rho}_1$ and $\Vec{\rho}_2$ in the hyperspherical coordinates as
\begin{equation}
\Vec{\rho} = \begin{pmatrix}
\sqrt{\frac{\mu_{12}}{\mu}}\Vec{\rho}_1\\
\sqrt{\frac{\mu_{3,12}}{\mu}}\Vec{\rho}_{2}
\end{pmatrix}, 
\end{equation}
where $\mu=m/\sqrt{3}$. This corresponds to Eq.~(88) in \cite{greene2017universal}. The corresponding hyperradial momentum conjugate to $\Vec{\rho}$ is given by
\begin{equation}
\Vec{P} = \begin{pmatrix}
\sqrt{\frac{\mu}{\mu_{12}}}\Vec{p}_1\\
\sqrt{\frac{\mu}{\mu_{3,12}}}\Vec{p}_{2}
\end{pmatrix}, 
\end{equation}
which corresponds to Eq.~(91) in \cite{greene2017universal}, such that $(\Vec{\rho},\Vec{P})$ form a conjugate pair which differs from the pairs given in \cite{greene2017universal} because of a typo,  
and the Hamiltonian in the hyperspherical coordinates is expressed as
 \begin{equation}
 H = \frac{\Vec{P}^2}{2\mu} + V(\Vec{\rho}\,).
 \end{equation}
 Here, the hyperradius is defined as $\mu\Vec{\rho}\,^2 = \mu_{12}\Vec{\rho_1}^2 + \mu_{3,12}\Vec{\rho_2}^2$. The Hamilton equations of motion in the hyperspherical coordinates are
 \begin{equation} \label{eom1}
     \frac{d \rho_i}{dt}=\frac{\partial H}{\partial P_i},
\end{equation}
\begin{equation} \label{eom2}
      \frac{d P_i}{dt}=-\frac{\partial H}{\partial \rho_i}.\\
 \end{equation}
We follow the same theoretical framework as Ref.~\cite{perez2014comparison} to set up the initial conditions, but do not align the 3D $z$-axis parallel to $\Vec{p}_2$. Using Avery's definition of hyperangles \cite{avery2012hyperspherical}, the initial hyperradial momentum vector $\Vec{P}_0$ can then be written as:
\begin{equation}
    \Vec{P}_0 =\begin{pmatrix*}[c]
         P_0\sin{\alpha_1^P}\sin{\alpha_2^P}\sin{\alpha_3^P}\sin{\alpha_4^P}\sin{\alpha_5^P}\\
        P_0\cos{\alpha_1^P}\sin{\alpha_2^P}\sin{\alpha_3^P}\sin{\alpha_4^P}\sin{\alpha_5^P}\\
        P_0\cos{\alpha_2^P}\sin{\alpha_3^P}\sin{\alpha_4^P}\sin{\alpha_5^P}\\
        P_0\cos{\alpha_3^P}\sin{\alpha_4^P}\sin{\alpha_5^P}\\
        P_0\cos{\alpha_4^P}\sin{\alpha_5^P}\\
        P_0\cos{\alpha_5^P}
    \end{pmatrix*},
\end{equation}
where $0\leq\alpha_1^P\leq2\pi$ and $0\leq\alpha_i^P\leq\pi$, $i=2,3,4,5$. The initial hyperradius $|\Vec{\rho}_0|=\rho_0$ is chosen such that the interactions are negligible, which fixes the magnitude of the initial momentum $P_0$ for a given energy as $E=P_0^2/2\mu$. The hyperangles $\alpha^P_i$ are generated using their respective probability distribution functions. The differential element for the $\alpha^P_i$'s is given as
\begin{align}
d\Omega_{P_0}=\sin^4{(\alpha^P_5)}\sin^3{(\alpha^P_4)}\sin^2{(\alpha^P_3)} \times\sin{(\alpha^P_2)}d\alpha^P_5d\alpha^P_4d\alpha^P_3d\alpha^P_2d\alpha^P_1.
\end{align}
The impact parameter vector $\Vec{b}$ is perpendicular to the initial momentum $\Vec{P}_0$ and is generated using the Gram-Schmidt orthogonalization. The parameters $\rho_0$, $\Vec{b}$, and $\Vec{P_0}$ are then sufficient to define the initial position vector $\Vec{\rho}_0=\Vec{b}-\sqrt{\rho_0^2-b^2}\Vec{P_0}/P_0$. With these initial conditions, the system is evolved using Eqs.~\ref{eom1} and ~\ref{eom2} to obtain the opacity function $\mathcal{P}(b,E)\{\Omega,\Delta\}.$ This is done by generating $n_{tra}=10^4$ trajectories for each impact parameter at a fixed collision energy $E$ and then counting the number of trajectories that recombine $n_{rec}(b,E)$. The recombination probability is then given by $\mathcal{P}(b,E)\{\Omega,\Delta\}=n_{rec}(b,E)/n_{tra}$. 

\section{S3. Incorporation of thermal dissociation}
The simplified picture above overestimates the loss rate from three-body recombination because a significant fraction of the molecules recombine into tetramers that are very weakly bound, and they can get collisionally dissociated almost immediately. The logic used to estimate the likelihood of a particular tetramer staying bound after it collides with a molecule from the thermal gas at temperature $T$ is as follows: consider the set of all recombined bound state energies $\{E_b(E)\}$ formed from an initial triad of molecules with relative collision energy $E$. For each tetramer binding energy, ${E_b(E)}_i$, $1000$ momenta $\{\Vec{P}_{th}\}$ are sampled from the Boltzmann distribution at temperature $T$. The momentum of the tetramer in the lab frame is given by $\Vec{P}_{rec}^{lab}=\Vec{P}_{rec}+\Vec{P}_{CM}$, where $\Vec{P}_{rec}^2/4m=(E+|{E_b}(E)_i|)/3$ and $\Vec{P}_{CM}$ is the momentum of the center of mass of the three initial colliding molecules in the lab frame, which is sampled from its Boltzmann distribution. We count the number of collisions $n_{dis}$ that have enough energy to dissociate the tetramer and replace the weight of each recombined event by $1-n_{dis}/1000$.  These weights are calculated for each recombination event ${E_b(E)}_i$ to obtain the modified opacity function used in further calculations $\Tilde{\mathcal{P}}(b,E,T)\{\Omega,\Delta\}$. 

\section{S4. Cross-section and rates}
The opacity function can be used to obtain the energy-dependent cross-section, $\sigma_3(E,T)$ for a sample temperature, and can be calculated via 
\begin{align}
\sigma_3(E,T) = \frac{8\pi^2}{3}\int_0^{\infty} \Tilde{\mathcal{P}}(b,E,T)\{ \Omega,\Delta \}b^4\,db,
\label{cross_section}
\end{align}
where $b$ is the impact parameter and $\Tilde{\mathcal{P}}(b,E,T)\{ \Omega,\Delta \}$ is the modified opacity function for a given set of microwave parameters $\{ \Omega,\Delta \}$. The temperature-dependent recombination rate, $K_3(T)$, can be obtained by averaging the energy-dependent three-body recombination rate, $k_3(E,T) =  \sigma_3(E,T) \sqrt{2E / \mu}$, over the Maxwell-Boltzmann distribution:
\begin{equation}
K_3(T) = \frac{1}{2(k_BT)^3}\int_0^{\infty} k_3(E,T)e^{-E/k_BT}E^2\,dE.
\label{recom_rate}
\end{equation}

\section{S5. Scaling laws}

\begin{figure}[t]
    \includegraphics[width=1\columnwidth]{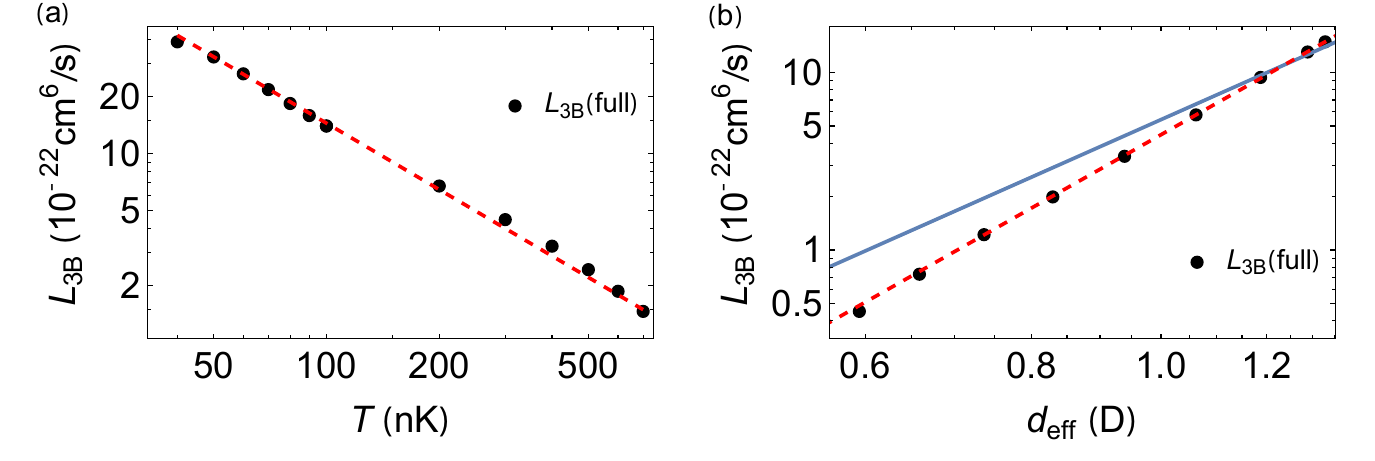}
    \caption{Scaling of the trajectory model. (a) A comparison of the calculated three-body loss coefficient, $L_{\mathrm{3B}}$(full) for $\Omega = 2 \pi \times 8$~MHz and $\Delta/\Omega = 1$ with the predicted scaling when no thermal dissociation is considered. The red dashed line shows a fit with $T^{-7/6}$, while the circles are results from the classical trajectory calculation. (b) Three body loss coefficient, $L_{\mathrm{3B}}$(full) for $\Omega = 2 \pi \times 8$~MHz shown as a function of effective dipole moment $d_{\mathrm{eff}}$. The circles are results from the classical trajectory calculation; the solid blue line is fit to the predicted scaling with $d_{\mathrm{eff}}^{10/3}$ and the red dashed is fit to $d_{\mathrm{eff}}^x$ with $x=4.2$.}
\label{fig:5}
\end{figure}

The unusual properties of dipolar recombination can be partially expected from scaling laws. The interaction energy roughly scales as ${d_{\rm eff}^2}/{r^3}$. The maximum impact parameter $b_\text{max}(E_k) = \sqrt{3}(d_{\rm eff}^2/2E_k)^{1/3}$ is defined as the distance where the interaction energy is equal to the collision energy $E_k$. Assuming that every trajectory with an impact parameter less than $b_\text{max}(E_k)$ recombines, the energy-dependent cross section scales as $\sigma_3(E_k) = (d_{\rm eff}^2 / 2E_k)^{5/3}$. Integrating over the Boltzmann distribution leads to $K_3(T)\propto T^{-7/6}d_{\rm eff}^{10/3}$. A calculation that does not include any thermal dissociation follows this temperature scaling quite well, as can be seen in Fig.~\ref{fig:5} (a). However, there is a deviation from the expected scaling with respect to the effective dipole moment $d_{\mathrm{eff}}$, shown in Fig.~\ref{fig:5} (b).

\section{S6. Kinetic model to fit experimental data}

In order to extract elastic and inelastic loss rates from lifetime data of the shielded molecular gases, we employ a fitting model that includes one-body, three-body, and evaporative losses. The following coupled differential equations describe the rate of change of molecule number and energy in the molecular gas:
\begin{align}
\dot{N} & = \dot{N}_\text{1B} + \dot{N}_\text{3B} + \dot{N}_\text{ev},  \\
\dot{E} & = \dot{E}_\text{1B} + \dot{E}_\text{3B} + \dot{E}_\text{ev}.
\end{align}
In our formalism, the total energy of the gas is $E = 3 N k_\mathrm{B} T$.  The one-body terms take the usual form $\dot{N}_\text{1B} = - N / \tau_\text{1B}$ and $\dot{E}_\text{1B} = - E / \tau_\text{1B}$, where $\tau_\text{1B}$ is the one-body lifetime. $\tau_\text{1B}$ is measured directly by observing low density loss curves in which other losses are negligible. The quantity $\tau_\text{1B}$ is kept fixed at the measured value. 

The three-body term in the number differential equation is given by \cite{olson2013optimizing} $ \dot{N}_\text{3B}  = -L_\text{3B} n_0^2 N / (3 \sqrt{3}) $. Here, $L_\text{3B}$ is the three-body loss rate coefficient and $n_0 = N \left(\bar{\omega}^2M/(2\pi k_B T)\right)^{3/2}$ is the peak density of the cloud, where $\bar{\omega} = (\omega_x \omega_y \omega_z)^{1/3}$ is the mean trap frequency, and $M$ the molecular mass. The three-body loss contribution to the energy differential equation is given by $ \dot{E}_\text{3B}  = - (2/3) L_\text{3B} n_0^2 E / (3 \sqrt{3}) + k_b T_h n_0^2 N / (3 \sqrt{3})$.  The $(2/3)$ prefactor comes from integrating the product of the energy density and the number density over the volume of the cloud. Finally, $T_h$ accounts for extra kinetic energy added during three-body loss, such as energy from the release of the binding energy~\cite{weber2003three}, secondary collisions from molecules escaping the trap, and from momentum correlations between the particles lost to recombination that cause deviations from the $(2/3)$ prefactor. 

The effects of evaporation are included via the term $\dot{N}_\text{ev} = -N \nu(\eta) \Gamma_{\rm el} / N_{\rm col} $  \cite{luiten1996kinetic, olson2013optimizing}. Here, $\nu(\eta)$ is the fraction of elastically scattered molecules with kinetic energy higher than the trap depth and $\Gamma_{\rm el} / N_{\rm col}$ is the thermalization rate.  $\Gamma_{\rm el}$ is the elastic scattering rate, $N_{\rm col}$ is the number of collisions to produce a $1/e$ change in the molecule temperature, $\eta = U_\text{min} / ( k_B T) $ is the truncation parameter and $U_\text{min}$ is the trap depth. From Ref.~\cite{davis1995analytical}, $\nu(\eta) = (2 + 2 \eta + \eta^2) / (2 e^{\eta})$.  The elastic scattering rate is $\Gamma_{\rm el} = \bar{n} \sigma_\text{el} v_{\rm th}$, where $\sigma_\text{el}$ is the elastic scattering cross-section, and $v_{\rm th} = 4\sqrt{k_B T / (\pi M)}$ is the thermal velocity. In our fitting routine, we cap $\Gamma_{\rm el} < \bar{\omega} / ( 2 \pi )$, in order to account for the hydrodynamic limit.  Because our gas is highly anisotropic, $N_{\rm col}$ is not a number, but rather a matrix accounting for the number of collisions for thermalization for every pair of trap axes, i.e.~$N_{\rm col}^{\rm xx}$, $N_{\rm col}^{\rm xy}$, $N_{\rm col}^{\rm xz}$, etc.  In our fitting routine, we fit the product $\sigma_\text{el} / N_{\rm col}$ and then use the calculated maximum $N_{\rm col}$ element to extract $\sigma_\text{el}$.

The evaporative term in the energy differential equation is $\dot{E}_\text{ev} = - (1/3) E \alpha(\eta) \Gamma_{\rm el} / N_{\rm col}$, where $\alpha(\eta) = (6 + 6\eta + 3\eta^2 + \eta^3) / (2 e^{\eta})$~\cite{davis1995analytical}.  $(1/3) E \alpha(\eta)$ is the energy of the molecules with kinetic energy larger than the trap depth and $\Gamma_{\rm el} / N_{\rm col}$ is the rate at which the energy will leave the system.

Experimentally, we measure the number and temperature of the molecular cloud as a function of hold time. Then the data is fitted with this model to extract the initial number, $N_0$, the initial temperature, $T_0$, and $L_{\rm 3B}$.  To obtain $\sigma_\text{el} / N_{\rm col}$, we first fit cross-thermalization data, see Ref.~\cite{bigagli2023collisionally}.

\section{S7. Dipole moment and dipolar energy}

\begin{figure}
    \centering
    \includegraphics[width = 8.6 cm]{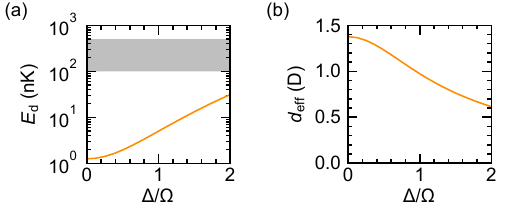}
    \caption{(a) Dipolar energy (orange line) compared to the experimentally accessible temperature range (grey shaded area) as a function of $\Delta / \Omega$. (b) Effective dipole moment as a function of $\Delta / \Omega$.}
    \label{fig:6}
\end{figure}

In Fig.~\ref{fig:6}, we show the dipolar energy and the induced dipole moment as a function of $\Delta / \Omega$, following equations reported in the main text. As can be seen from Fig.~\ref{fig:6} (a), all our experimental data is taken in the regime $T \gg E_{\rm d}$, justifying a semi-classical treatment.

